%% file: letterG_A.tex
\documentclass[twocolumn,10pt,a4paper,superscriptaddress,showpacs,preprintnumbers,amsmath,amssymb,notitlepage,nofootinbib]{revtex4-1}
\date{March 2010}
\usepackage{epsfig}
\usepackage{latexsym}
\usepackage{amssymb}
\usepackage{booktabs}
\usepackage{graphicx}
\newcommand{\be}{\begin{equation}}
\newcommand{\ee}{\end{equation}}
\newcommand{\ba}{\begin{eqnarray}}
\newcommand{\ea}{\end{eqnarray}}
\newcommand{\bi}{\begin{itemize}}
\newcommand{\ei}{\end{itemize}}

\newcommand{\<}{\langle}
\renewcommand{\>}{\rangle}

\newcommand{\la}{\label}

\newcommand{\txts}{\textstyle}

\newcommand{\mev}{\textrm{MeV}}
\newcommand{\gev}{\textrm{GeV}}
\newcommand{\fm}{\textrm{fm}}

\begin{document}
\preprint{MITP-22-053}
\title{The isovector axial form factor of the nucleon from lattice QCD}

\author{Dalibor~Djukanovic} 
\affiliation{Helmholtz-Institut Mainz, Johannes Gutenberg-Universit\"at Mainz,
D-55099 Mainz, Germany}
 \affiliation{GSI Helmholtzzentrum für Schwerionenforschung, Darmstadt (Germany)}

\author{Georg~von~Hippel}
\affiliation{PRISMA$^+$ Cluster of Excellence \& Institut f\"ur Kernphysik,
 Johannes Gutenberg-Universit\"at  Mainz,  D-55099 Mainz, Germany}

\author{Jonna~Koponen}
\affiliation{PRISMA$^+$ Cluster of Excellence  \& Institut f\"ur Kernphysik,
Johannes Gutenberg-Universit\"at Mainz, D-55099 Mainz, Germany}

\author{Harvey~B.~Meyer} 
\affiliation{Helmholtz-Institut Mainz, Johannes Gutenberg-Universit\"at Mainz,
D-55099 Mainz, Germany}
 \affiliation{PRISMA$^+$ Cluster of Excellence  \& Institut f\"ur Kernphysik,
Johannes Gutenberg-Universit\"at Mainz, D-55099 Mainz, Germany}

\author{Konstantin~Ottnad}
\affiliation{PRISMA$^+$ Cluster of Excellence \& Institut f\"ur Kernphysik,
 Johannes Gutenberg-Universit\"at  Mainz,  D-55099 Mainz, Germany}

\author{Tobias~Schulz}
\affiliation{PRISMA$^+$ Cluster of Excellence \& Institut f\"ur Kernphysik,
 Johannes Gutenberg-Universit\"at  Mainz,  D-55099 Mainz, Germany}

\author{Hartmut~Wittig} 
\affiliation{Helmholtz-Institut Mainz, Johannes Gutenberg-Universit\"at Mainz,
D-55099 Mainz, Germany}
 \affiliation{PRISMA$^+$ Cluster of Excellence  \& Institut f\"ur Kernphysik,
Johannes Gutenberg-Universit\"at Mainz, D-55099 Mainz, Germany}

 \email{meyerh@uni-mainz.de}

\date{\today}

\begin{abstract}
The isovector axial form factor of the nucleon plays a key role in
interpreting data from long-baseline neutrino oscillation experiments.
We perform a lattice-QCD based calculation of this form factor,
introducing a new method to directly extract its $z$-expansion from
lattice correlators.  Our final parametrization of the form factor,
which extends up to spacelike virtualities of $0.7\,{\rm GeV}^2$ 
with fully quantified uncertainties, agrees with previous lattice
calculations but is significantly less steep than neutrino-deuterium
scattering data suggests.
\end{abstract}

\maketitle

\section{Introduction}

The axial form factor of the nucleon $G_{\rm A}(Q^2)$ plays a central role in understanding
the quasi-elastic part of GeV-scale neutrino-nucleus cross sections.
Particularly for the upcoming long-baseline neutrino oscillation experiments DUNE~\cite{DUNE:2015lol} and T2HK~\cite{Hyper-Kamiokande:2018ofw},
these cross sections must be known with few-percent uncertainties~\cite{Ruso:2022qes}
to enable a sufficiently reliable reconstruction of the incident neutrino energy.
In the absence of modern, high-quality experimental measurements of $G_{\rm A}(Q^2)$~\cite{Bernard:2001rs}, calculations for the axial
form factor from lattice QCD~\cite{Meyer:2022mix} are of crucial importance  in order to
maximize the scientific output of neutrino-oscillation experiments.

For a long time, the axial charge of the nucleon, $G_{\rm A}(0)$,
served as a benchmark quantity for lattice QCD
calculations~\cite{Aoki:2021kgd}, exemplifying the improvements of
recent years in terms of control over statistical and systematic
errors. The latter are caused mainly by the excited-state
contamination in Euclidean correlation functions, as well as by the
chiral and continuum extrapolation.  Many of the techniques developed
have been carried over and applied to non-vanishing momentum transfer
$Q^2$, most
recently in Refs.~\cite{Park:2021ypf,Jang:2019vkm,RQCD:2019jai,Alexandrou:2020okk,Shintani:2018ozy,Ishikawa:2021eut}.
In comparison to the calculation of the charge, a new source of
systematics arises for the form factor, namely the parameterization of the
$Q^2$-dependence. Historically, an \emph{ad hoc} dipole ansatz was used
(see~\cite{Bernard:2001rs}), incurring an unquantified
model systematic. As a modern alternative, an ansatz based on the
$z$-expansion has been used extensively, leading to less model bias
at the cost of an increased statistical error on the phenomenological
determination of the mean square radius
$\<r_{\rm A}^2\> = [\frac{-6}{G_{\rm A}} \frac{d G_{\rm A}}{dQ^2}]_{Q^2=0} $~\cite{Hill:2017wgb}.
The sensitivity to the parameterization is also visible in lattice
calculations, where the different ans\"atze lead to inconsistent
results (see e.g.\ \cite{RQCD:2019jai}).

In this Letter, we perform a high-statistics calculation of $G_{\rm
  A}(Q^2)$ for momentum transfers up to $0.7\,{\rm GeV}^2$ using
lattice simulations with dynamical up, down and strange quarks with an
O($a$) improved Wilson fermion action. We employ a new analysis method
that simultaneously handles the issues of the excited-state contamination
and the description of the form factor's $Q^2$ dependence.

\section{Methodology}
\label{sec:methodology}

The matrix elements of the local iso-vector axial current
$A^{a}_{\mu}(x) = \bar\psi \gamma_\mu\gamma_5\frac{\tau^a}{2}\psi$
between single-nucleon states are  parameterized by
the axial form factor $G_{\rm A}(Q^2)$ and induced pseudoscalar
form factor $G_{\rm P}(Q^2)$. 
We focus on the current component orthogonal to the momentum transfer, thereby projecting out the axial form factor,
\begin{eqnarray}\label{Eq:Amu_mat}
  && \< N(p^{\prime},s^\prime)|\, \vec q\times \vec A^{a}(0) \,|N(p,s)\> =
  \\ \nonumber &&
\qquad  G_{\rm A}(Q^2) \;\bar{U}^{s'}(p^{\prime})  \,\vec q\times \vec\gamma\; \gamma_5  \,{\txts\frac{\tau^a}{2}} U^{s}(p),
\end{eqnarray}
where $\vec q=\vec p^{\,\prime}-\vec p$, $Q^2 = \vec q{\,}^2 - (E_{\vec p'} - E_{\vec p})^2$
and $U^s(p)$ is an isodoublet Dirac spinor with momentum $p$ and spin state $s$.
We employ Euclidean notation throughout.

The setup for our lattice determination of the  axial form factor is 
very similar to the one we used in the case of the electromagnetic
form factors \cite{Djukanovic:2021cgp}. 
The nucleon two- and three-point functions are computed as
\begin{eqnarray}
\label{Eq:C2pt}
C_2(\vec{p},t) &=& a^3\sum_{\vec{x}} e^{i\vec{p}\cdot\vec{x}}\, \Gamma_{\beta\alpha}\,\Big\langle \Psi^\alpha(t,\vec{x})\overline{\Psi}^\beta(0)\Big\rangle,
\\
 C_{3}(\vec{q},t,t_s) &=& -i\,a^6\sum_{\vec{x},\vec{y}}
e^{i\vec{q}\cdot\vec{y}}\,\Gamma_{\beta\alpha}\, \frac{\vec q\times \vec s}{|\vec q\times \vec s|^2}\cdot 
\la{Eq:C3pt}
\\ && \Big\langle \Psi^\alpha(t_s,\vec{x}) \,
 \vec q\times\! \vec A^{a=3}(t,\vec{y})\,\overline{\Psi}^\beta(0)\Big\rangle,
\nonumber
\end{eqnarray}
where $\Psi^\alpha(\vec{x},t)$ denotes the proton interpolating
operator 
\begin{equation}
\Psi^\alpha(x) =  \epsilon_{abc} \big( \tilde{u}^T_a(x) C \gamma_5 \tilde{d}_b(x) \big) \tilde{u}^\alpha_c(x).
\end{equation}
The quark fields $\tilde u,\tilde d$ are  smeared with a Gaussian kernel~\cite{Gusken:1989ad},
using  APE-smeared gauge fields~\cite{Albanese:1987ds}.
Note that the nucleon three-point function is computed in the rest frame
of the final-state nucleon,  $\vec{p}^{\,\prime}=0$, and
the chosen projection matrix $\Gamma$ reads
\begin{equation}
\Gamma = {\txts\frac{1}{2}}(1 + \gamma_0)(1 + i \gamma_5 \vec s\cdot \vec\gamma).
\end{equation}
In practice, we have set $\vec s = \vec e_3$, i.e. the nucleon spin is aligned
along the $x_3$-axis. When averaging over equivalent momenta we find an
improved signal using the constraint $|q_3|\leq\min \Bigl(|q_1|,|q_2|\Bigr)$.
The transverse part $\vec q\times\! \vec A^{a}$ of the axial current receives no additive O($a$) improvement.
For its multiplicative renormalization, we employ the determination of $Z_A$ and $b_A$
from~\cite{DallaBrida:2018tpn} and~\cite{Korcyl:2016ugy}, respectively, while the coefficient $\tilde b_A$
in the notation of~\cite{Korcyl:2016ugy} is neglected, since it parametrizes a sea-quark effect
and is expected to be small.

We use the ratio
\begin{equation}
\label{Eq:ratio}
 R(\vec{q},t,t_s) \equiv \frac{C_{3}(\vec{q},t,t_s)}{C_2(0,t_s)} \sqrt{ \frac{C_2(\vec{q},t_s-t)C_2(\vec{0},t)C_2(\vec{0},t_s)}{C_2(\vec{0},t_s-t)C_2(\vec{q},t)C_2(\vec{q},t_s)}}
\end{equation}
to build the summed insertion
\ba
\label{eq:summation}
S(\vec q,t_s) &\equiv&  a\, \sqrt{\frac{2 E_{\vec q}}{m+E_{\vec q}}}\;\sum_{t=a}^{t_s-a}  R(\vec{q},t,t_s)
\\ &\stackrel{t_s\to\infty}{=}& b_0(\vec q) + t_s   G_{\rm A}(Q^2)    
+ \dots
\nonumber
\ea
The dots stand for excited-state contributions that are of order $t_s e^{-\Delta t_s}$,
with $\Delta$ the energy gap above the single-nucleon state.
As a novelty, we introduce a technique which is based on fitting the quantities $S(\vec q,t_s)$
simultaneously for different $\vec q$ and $t_s$,
by parameterizing the axial form factor from the outset via the $z$-expansion (see~\cite{Hill:2017wgb,Hill:2010yb} and Refs.\ therein),
\begin{eqnarray}\label{Eq:zexp}
  G_{\rm A}(Q^2) &=& \sum^{n_{\rm max}}_{n=0} a_n \,z^n(Q^2) \,,
  \\
  z(Q^2)  &=& \frac{\sqrt{t_{\rm cut} + Q^2} - \sqrt{t_{\rm cut}} }{\sqrt{t_{\rm cut} + Q^2} + \sqrt{t_{\rm cut}}}\;.
  \la{eq:zvar}
\end{eqnarray}
The fit parameters are the coefficients $a_n$ and 
the offsets $b_0(\vec q)$, which we keep as independent fit
parameters for each $\vec q$.
In the data analyzed below,  we set $n_{\rm max} =2$
without constraining the fit parameters by priors. We note that setting $n_{\rm max} =3$
would require the use of priors for the highest-order term to stabilize the fit,
but the results are consistent with our preferred $n_{\rm max} =2$ results.
To obtain the form factor at the physical point, the $a_n$ are extrapolated to the continuum
and interpolated to the physical pion mass, at which point the form factor may be evaluated
at any virtuality in the chosen expansion interval $[0,\;0.7\,{\rm GeV}^2]$.

Our method relies on the fact that, for a given $Q^2$-interval, the $z$-expansion represents
a general, systematically improvable parameterization of the form
factor~\cite{Hill:2010yb}. We have chosen to map $Q^2=0$ to the point $z=0$
and  set $t_{\rm cut} =  (3M_\pi^{\rm phys})^2$ to the three-pion kinematic
threshold at the physical pion mass for all gauge ensembles
used, as this choice facilitates the chiral extrapolation of the $a_n$.
We find that the immediate parameterization of the form factor has a
stabilizing effect as compared to the standard two-step procedure of first
obtaining the form factor independently at discrete values of $\vec
q$, followed by a continuous parameterization of these data points.

\begin{figure}[!t]    
\centering
\includegraphics[width=0.8\columnwidth]{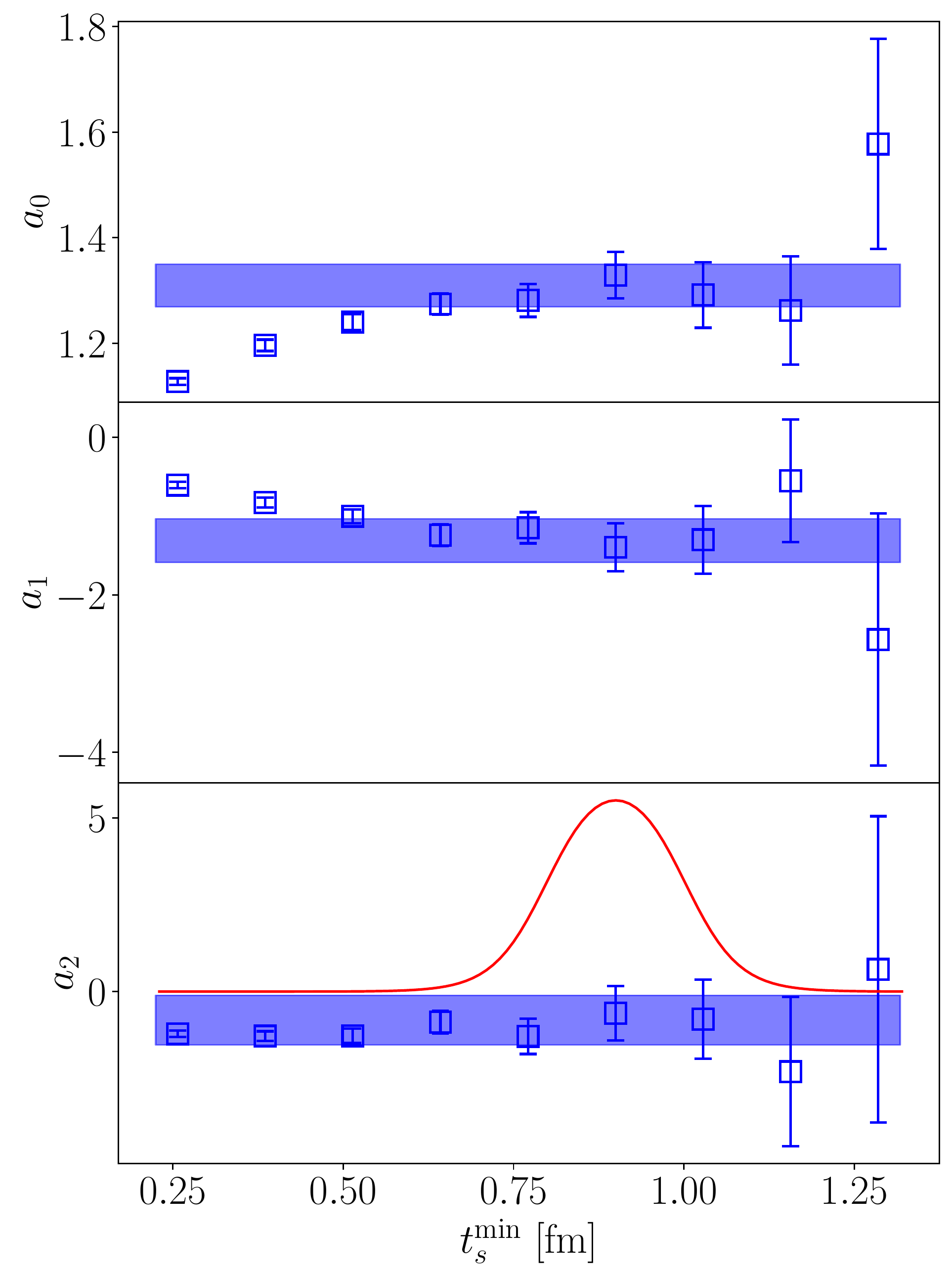}
\caption{Illustration of averaging over the minimum source-sink separation $t_{s}^{\mathrm{min}}$ in the summation method for the near-physical pion mass ensemble E250 of size $192\times96^3$ with a lattice spacing of 0.064\,fm~\cite{Bruno:2016plf}. We perform the $z$-expansion fits for each ensemble starting at different values of $t_{s}^{\mathrm{min}}$. The results for coefficients $a_0$, $a_1$ and $a_2$ are shown here as the blue squares. The bands represent the smooth-window averages over $t_{s}^{\mathrm{min}}$, and the solid red line shows the weight function Eq.~\eqref{eq:smoothwindow} (multiplied by ten for clarity).}
\label{fig:tsep_min_average_E250}
\end{figure}

We perform fits to $S(\vec q,t_s)$ based on the second line of Eq.\ (\ref{eq:summation}), dropping
the omitted terms, and including all values of $t_s$ greater than or equal to
a certain $t_s^{\textrm{min}}$.
At small values of $t_s$, contributions from excited states are expected to be significant, whereas at
large $t_s$ the signal-to-noise ratio becomes poor.
This leaves us with a relatively small window of starting values $t_s^{\textrm{min}}$ that can safely be used.
Rather than choosing a single $t_s^{\textrm{min}}$, we average the fit results over $t_s^{\textrm{min}}$
using as a weight factor the `smooth window' function
\begin{equation}
  \frac{1}{{\cal N}_w}
  \left[\tanh{\left(\frac{t_s^{\textrm{min}}-t_w^{\textrm{low}}}{\Delta t_w}\right)}-
  \tanh{\left(\frac{t_s^{\textrm{min}}-t_w^{\textrm{up}}}{\Delta t_w}\right)}\right]
\label{eq:smoothwindow}  
\end{equation}
with $t_w^{\textrm{low}}=0.8$~fm, $t_w^{\textrm{up}}=1.0$\;fm and $\Delta t_w=0.08$\;fm.
The weights are normalized by ${\cal N}_w$ so as to add up to unity.
The average represents very well what could be identified as
a plateau in the fit results, as illustrated in Fig.~\ref{fig:tsep_min_average_E250}.
The three panels also illustrate
the advantage of having to scrutinize only very few observables for excited-state effects, as opposed
to having to do this for every $Q^2$ value.
Having an extended set of $t_s$ values at our disposal, the control over these effects is significantly improved
as compared to our previous summation-method results for the vector form factors~\cite{Djukanovic:2021cgp}.

\section{The lattice calculation}
\label{sec:lattice}

We use a set of fourteen CLS $N_f=2+1$ ensembles~\cite{Bruno:2014jqa} that have been generated 
with non-perturbatively $\mathcal{O}(a)$-improved Wilson
fermions~\cite{Sheikholeslami:1985ij,Bulava:2013cta}
and the tree-level improved L\"uscher-Weisz gauge action~\cite{Luscher:1984xn}.
They cover the range of lattice spacings from $0.050$\;fm to $0.086$\;fm and pion masses
from about $350~\mev$ down to $130~\mev$. 
For most of these ensembles, the fields obey open boundary conditions in time~\cite{Luscher:2011kk}
in order to prevent topological freezing~\cite{Schaefer:2010hu}.
The reweighting factors needed to correct for the treatment of the
strange-quark determinant during the gauge field generation were computed
using the method of Ref.~\cite{Mohler:2020txx}.
Our setup to compute the nucleon two- and three-point functions
is similar to that used in our study on the isovector charges
of the nucleon~\cite{Harris:2019bih}.

As discussed in section~\ref{sec:methodology}, we perform simultaneous fits to all data points with
$Q^2\leq 0.7~\gev^2$
and source-sink separations $t_s \geq t_s^{\textrm{min}}$ on each ensemble to
obtain the coefficients $a_i$ of the $z$-expansion at the given pion mass, lattice spacing
and volume. Ensemble-by-ensemble results are compiled in
the Supplementary Material. We then proceed to perform chiral and continuum
extrapolations of the coefficients $a_i$ to the physical point,
including for each of them a term linear in $a^2$.
As for their chiral behaviour, we use the following three ans\"atze:
\begin{enumerate}
\item
  Linear in $M_\pi^2$ for all coefficients $a_i$.
  
\item
  Again linear in $M_\pi^2$ for coefficients $a_1$ and $a_2$,
  and an extended ansatz containing a chiral logarithm for the zeroth coefficient:
  \begin{equation*}
    a_0 = g_a^{(0)}+g_a^{(1)}M_{\pi}^2+g_a^{(3)}M_{\pi}^3-g_a^{(2)}M_{\pi}^2\ln{\frac{M_{\pi}}{M_n}}
  \end{equation*}
  with
  \begin{align*}
    g_a^{(1)} &= 4d_{16}-\frac{(g_a^{(0)})^3}{16\pi^2 F_\pi^2},\\
    g_a^{(2)} &= \frac{g_a^{(0)}}{8\pi^2 F_\pi^2}\left(1+2(g_a^{(0)})^2\right),\\
    g_a^{(3)} &= \frac{g_a^{(0)}}{8\pi F_\pi^2 M_n}\left(1+(g_a^{(0)})^2\right)-\frac{g_a^{(0)}}{6\pi F_\pi^2}\Delta_{c_3,c_4},
  \end{align*}
  where  $M_n=938.92~\mev$ is the nucleon mass and  $F_\pi=92.42~\mev$ the pion decay constant~\cite{Schindler:2006it}.
  Here $\Delta_{c_3,c_4}=c_3-2c_4$ is a combination of low-energy constants $c_3$ and $c_4$. The free fit parameters for
  the zeroth coefficient's chiral extrapolation are $g_a^{(0)}$, $d_{16}$ and $\Delta_{c_3,c_4}$.
  
\item
  Same as ansatz~2, but including $M_\pi^3$ terms for coefficients $a_1$ and $a_2$.
  
\end{enumerate}
Note that, while the coefficients $a_i$ do not have common fit parameters, they are correlated
within an ensemble: these correlations are taken into account in the fits.
If the resulting correlation matrix is larger than $70\times 70$, we damp the off-diagonal
correlations by 0.5\%\ldots1.5\% to avoid numerical instabilities~\cite{Touloumis_2015}.

We perform multiple extrapolations  using the different fit ans\"atze described above with
pion mass cuts $M_{\pi} < M_{\pi}^{\rm cut}$ with $M_{\pi}^{\rm cut} [{\rm MeV}]\in\{ 300,\,285,\,265,\,250\}$,
as well as dropping data from the coarsest lattice spacing, to get a handle on systematic effects.
Although we do not observe a strong dependence on the volume, we include a term~\cite{Beane:2004rf}
\begin{equation}
  \frac{M_{\pi}^2}{\sqrt{M_{\pi}L}}\mathrm{e}^{-M_{\pi}L}
  \label{eq:FSE}
\end{equation}
for the zeroth coefficient $a_0$ to check for possible finite-size effects (FSE) in some of the
extrapolation fits.
For a subset of fits, we impose Gaussian priors on the coefficients
multiplying the $a^2$ terms, restricting the difference between the
values at the coarsest lattice spacing and in the continuum limit to at
most 20\%. This is motivated by a tendency of these fits to attribute
unnaturally large corrections to discretization effects,
especially for $a_1$ and $a_2$ that are statistically less precise.
We keep those fits that have a $p\,$-value better than 5\% and
provide a satisfactory description of the data, especially at pion masses below 200\,MeV.

Some examples of these fits based on different ans\"atze and pion mass cuts are
shown in the Supplementary Material.
While most of our fits have a good $p\,$-value without including the FSE term of Eq.\ (\ref{eq:FSE}),
which tends to slightly increase the uncertainties, 
we do  include these fits in the analysis in order to account for the systematic effect due to finite-size
corrections. We can also inspect finite-size effects directly by comparing our results
of the $z$-expansion fits on two ensembles at a pion mass of $280\,\mev$, H105 and N101, which differ only by their spatial sizes,
$L=2.8\,$fm and 4.1\,fm respectively.
We find that the coefficients $a_i$ agree well, confirming that finite-size effects are small
at the current level of precision.

Since different fit ans\"atze and cuts can be equally well motivated,
as in our previous study of the vector form factors of the nucleon~\cite{Djukanovic:2021cgp}
we perform a weighted average~\cite{Jay:2020jkz} over the resulting $a_i$, where the Akaike Information
Criterion (AIC)~\cite{Akaike} is used to weight different analyses and to estimate
the systematic error associated with the variations of the global fit. Different versions of
the AIC weights have been developed and used over the years. Here we choose~\cite{Borsanyi:2020mff}
\begin{equation}
  w^{\textrm{AIC}}=N\mathrm{e}^{-\frac{1}{2}\left(\chi^2+2n_{\textrm{par}}-n_{\textrm{data}}\right)},
\end{equation}
where the minimum  $\chi^2$, the number of fit parameters $n_{\textrm{par}}$ and the number of data points
$n_{\textrm{data}}$ characterize the fit.
$N$ is a normalization factor that ensures that the sum of the weights is unity.
The corresponding cumulative distribution functions of the coefficients $a_i$ and of
the mean square radius $\langle r_A^2\rangle$ are well-behaved and show no outliers.
We determine the central value from the 50th percentile, and the full uncertainty
as the interval from the 16th to the 84th percentile. The decomposition of the  error
into its statistical and systematic components is achieved  following the prescription proposed in~\cite{Borsanyi:2020mff}.

\section{Results}

Our results  for the coefficients of the $z$-expansion of the nucleon axial form
factor in the continuum and at the physical pion mass are
\begin{align}
  a_0 =& \quad\: 1.225 \pm 0.039 \textrm{ (stat)} \pm 0.025 \textrm{ (syst)}, \nonumber \\
  a_1 =& -1.274 \pm 0.237 \textrm{ (stat)} \pm 0.070 \textrm{ (syst)}, \nonumber \\
  a_2 =& -0.379 \pm 0.592 \textrm{ (stat)} \pm 0.179 \textrm{ (syst)} 
\label{eq:results_ai}  
\end{align}  
with a correlation matrix
\begin{equation}
M_{\textrm{corr}} = \left(\!
\begin{array}{rrr}
 1.00000 & -0.67758 &  0.61681 \\
-0.67758 &  1.00000 & -0.91219 \\
 0.61681 & -0.91219 &  1.00000
\end{array} \,\right).
\end{equation}
These results, meant to be inserted into Eqs.\ (\ref{Eq:zexp}--\ref{eq:zvar}) with $t_{\rm cut}=(3M_{\pi^0})^2$,
lead to the following mean square radius,
\ba
\langle r^2_A \rangle &=&
(0.370 \pm 0.063 \textrm{ (stat)} \pm 0.016 \textrm{ (syst)} )~\fm^2.~~~~
\ea

\begin{figure}[!t]    
\centering
\includegraphics[width=0.98\columnwidth]{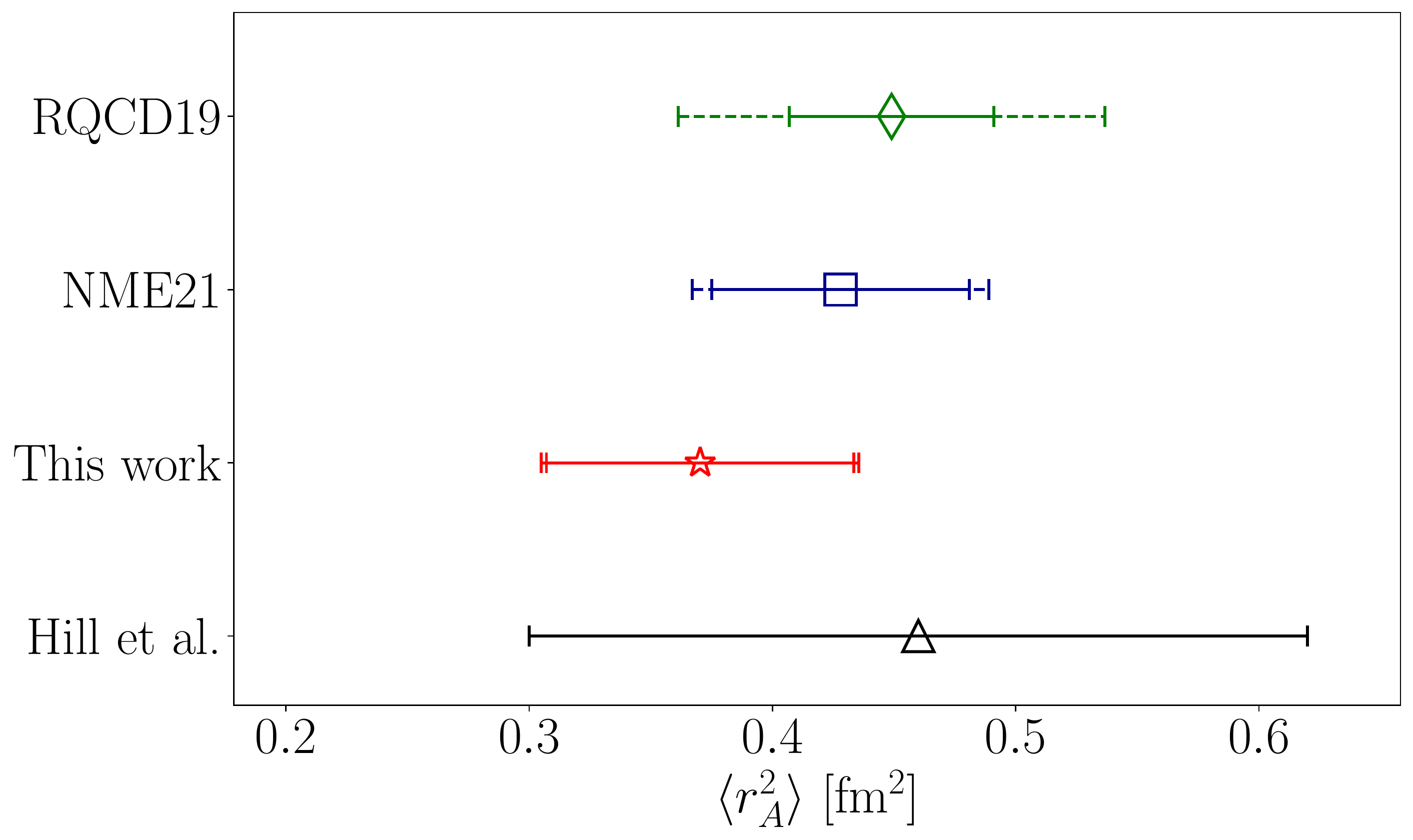}
\caption{Comparison of lattice determinations of the mean square radius
$\langle r_\text{A}^2\rangle$, from Refs.~\cite{Park:2021ypf}
(NME21) and \cite{RQCD:2019jai} (RQCD19). The point labeled
Hill et al. is  an average of the values obtained from $z$-expansion
fits to neutrino scattering and muon capture \cite{Hill:2017wgb}.
The smaller error bars with solid lines show the statistical errors, whereas
the wider error bars with dashed lines show the total errors (including systematic
uncertainties).}
\label{fig:rAsq_comparison}
\end{figure}

We compare our result to other lattice QCD determinations of the mean square radius in
Fig.~\ref{fig:rAsq_comparison}, finding good agreement. The comparison features only
lattice calculations with a full error budget, including a continuum extrapolation;
see Refs.\ \cite{Jang:2019vkm,Alexandrou:2020okk,Shintani:2018ozy,Ishikawa:2021eut,Hasan:2017wwt} for further lattice results.
The NME21 result is from~\cite{Park:2021ypf}, and the RQCD19 result is from~\cite{RQCD:2019jai}.
Both studies parameterize the $Q^2$ dependence of the form factor using a $z$-expansion (RQCD
also use a dipole ansatz as an alternative parameterization, but that result is not shown in
the figure). For comparison, we show the average of the values obtained from $z$-expansion
fits to neutrino scattering and muon capture measurements~\cite{Hill:2017wgb}. Our result also agrees well
with the earlier two-flavour calculation by the Mainz group~\cite{Capitani:2017qpc}, and with
a more recent analysis~\cite{Schulz:2021kwz} by the same group that has
been obtained via the conventional two-step process of first determining the form factor at discrete
$Q^2$ values and subsequently parameterizing it.

\begin{figure}[!t]    
\centering
\includegraphics[width=0.98\columnwidth]{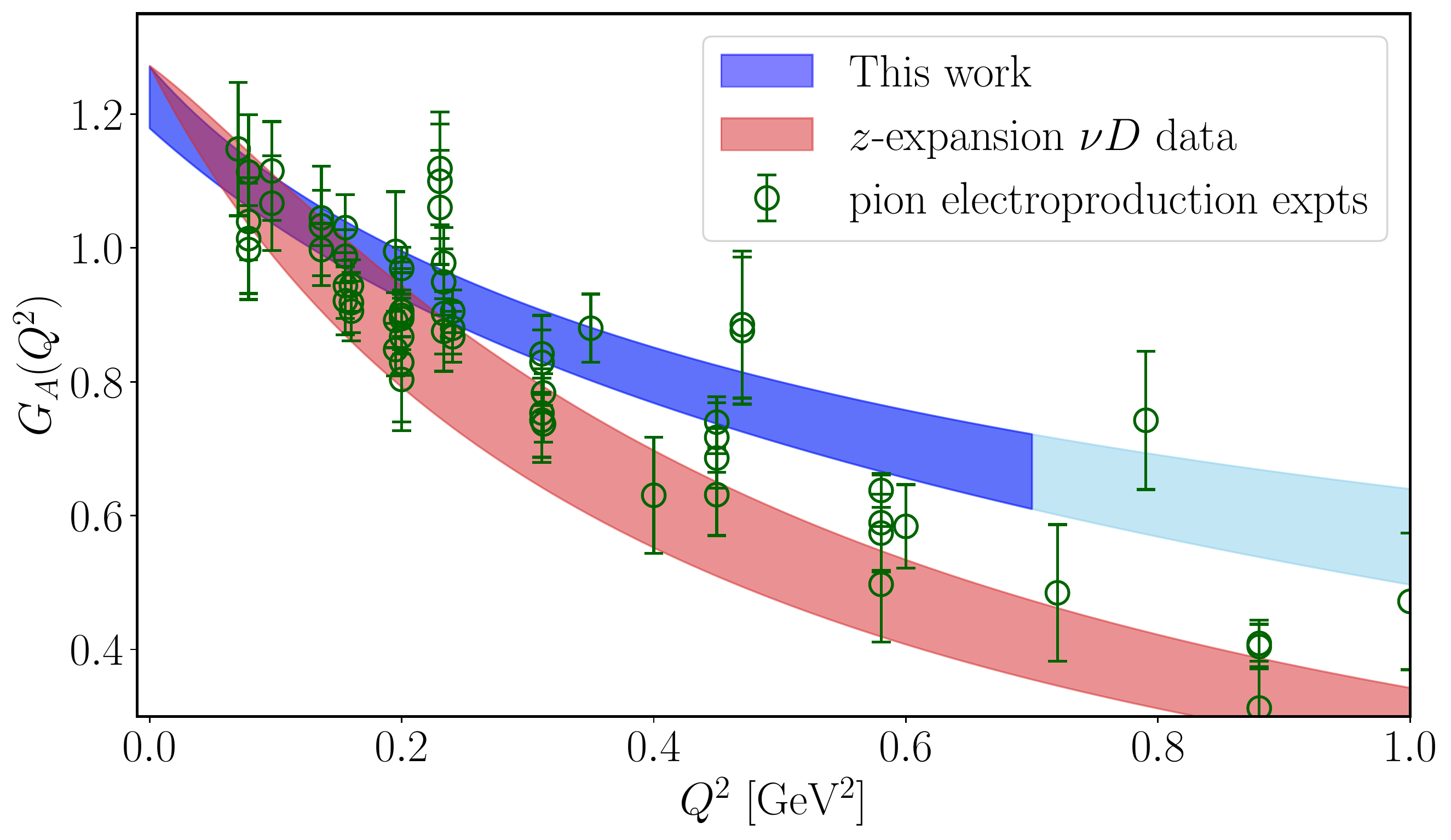}
\caption{Comparing our result for the axial form factor to data from pion electroproduction
  experiments~\cite{Bernard:2001rs} and to a $z$-expansion fit to neutrino-Deuterium scattering
  data~\cite{Meyer:2016oeg}. There is a clear tension between the lattice QCD result and the
  $z$-expansion extracted from deuterium bubble chamber data, especially at large $Q^2$. The
  darker blue error band highlights the $Q^2$ range of our lattice data. The data from
  electroproduction experiments has been multiplied by the current PDG value for the axial charge~\cite{PDG2022}.}
\label{fig:physFF_and_expt}
\end{figure}

Perhaps even more interesting is the comparison of our result
 for the axial form factor to data
from pion electroproduction experiments~\cite{Bernard:2001rs} and to a $z$-expansion fit
to neutrino-Deuterium scattering data~\cite{Meyer:2016oeg} in Fig.~\ref{fig:physFF_and_expt}.
Our result agrees well with other lattice QCD calculations, as can be seen by comparing
this figure to Fig.~3 in the review~\cite{Meyer:2022mix}, but there is a tension with the
axial form factor extracted from experimental deuterium bubble chamber data~\cite{Meyer:2016oeg}.
This tension is strongest at large $Q^2$, the deuterium extraction being lower than the
lattice prediction. The authors of the Snowmass White Paper on
Neutrino Scattering Measurements~\cite{Alvarez-Ruso:2022ctb} remark that, when translated to the
nucleon quasielastic cross section, this discrepancy suggests that a 30-40\% increase would be needed
for these two results to match. They also note that recent high-statistics data on nuclear targets
cannot directly resolve such discrepancies due to nuclear modeling uncertainties, and that new
elementary target neutrino data would provide a critical input to resolve such discrepancies.
  
\section{Conclusions}

In this Letter we have introduced a new method to extract the axial form factor of the nucleon.
It combines two well-known methods into one analysis step:
the summation method ensures that excited-state effects
are sufficiently suppressed, and the $z$-expansion  readily provides the parameterization of
the $Q^2$ dependence of the form factor. Our main results are the coefficients of the
$z$-expansion, given in Eq.~(\ref{eq:results_ai}).
Systematic effects are included through AIC averaging, which also
provides the break-up into statistical and systematic uncertainties and
the correlations among the coefficients. Our results are
statistics-limited, implying that significant improvements are still
straightforwardly possible, though computationally costly.

We observe good agreement with other lattice QCD determinations of the
axial form factor, which means that the tension with the shape of the
form factor extracted from deuterium bubble chamber data is further
strengthened. Comparing our result for $a_0\equiv G_{\rm A}(0)$  to the Particle Data
Group (PDG) value for the axial charge, $g_A = 1.2754(13)$ \cite{PDG2022},
which one might view as a benchmark, we find agreement at the
$1.1\,\sigma$ level. Also, using largely the same gauge ensembles as
in this work, we have previously found a good overall agreement for
the isovector vector form factors~\cite{Djukanovic:2021cgp} with
phenomenological determinations, which are far more precise than in
the axial-vector case. Thus a nucleon axial form factor falling off less steeply 
than previously thought now appears more likely.

In the near future, we plan to perform a dedicated calculation of
various forward nucleon matrix elements, including the axial charge,
updating the results of Ref.~\cite{Harris:2019bih}.

\acknowledgments{We thank Tim Harris, who was involved in the early
  stages of this project~\cite{Brandt:2017vgl}. This work was
  supported in part by the European Research Council (ERC) under the
  European Union’s Horizon 2020 research and innovation program
  through Grant Agreement No.\ 771971-SIMDAMA and by the Deutsche
  Forschungsgemeinschaft (DFG) through the Collaborative Research
  Center SFB~1044 ``The low-energy frontier of the Standard Model'',
  under grant HI~2048/1-2 (Project No.\ 399400745) and in the Cluster
  of Excellence “Precision Physics, Fundamental Interactions and
  Structure of Matter” (PRISMA+ EXC 2118/1) funded by the DFG within
  the German Excellence strategy (Project ID 39083149).
  Calculations for this project were partly performed on the HPC
  clusters ``Clover'' and ``HIMster2'' at the Helmholtz Institute Mainz,
  and ``Mogon 2'' at Johannes Gutenberg-Universit\"at Mainz.
  The authors gratefully acknowledge the Gauss Centre for Supercomputing e.V. (www.gauss-centre.eu) for funding this project by providing computing time on the GCS Supercomputer systems JUQUEEN and JUWELS at J\"ulich Supercomputing Centre (JSC) via grants HMZ21, HMZ23 and HMZ36 (the latter through the John von Neumann Institute for Computing (NIC)), as well as on the GCS Supercomputer HAZELHEN at H\"ochstleistungsrechenzentrum Stuttgart (www.hlrs.de) under project GCS-HQCD.
  
Our programs use the QDP++ library~\cite{Edwards:2004sx} and deflated SAP+GCR
solver from the openQCD package~\cite{Luscher:2012av}, while the contractions
have been explicitly checked using~\cite{Djukanovic:2016spv}. We are grateful to
our colleagues in the CLS initiative for sharing the gauge field configurations
on which this work is based.
}

 \bibliography{GAref.bib} 

\newpage
 
\include{suppl}

\end{document}

%% file: suppl.tex
\twocolumngrid

\section*{Supplementary material} 

In order to further document our results, we provide additional
figures illustrating a few specific aspects of the analysis, and
expand on certain technical details.

\subsection{Lattice ensembles}

As explained in the main text, we use the CLS $N_f=2+1$ ensembles \cite{Bruno:2014jqa} that have been generated 
with non-perturbatively $\mathcal{O}(a)$-improved Wilson fermions \cite{Sheikholeslami:1985ij,Bulava:2013cta}
and the tree-level improved L\"uscher-Weisz gauge action \cite{Luscher:1984xn}. The lattice spacings of these
ensembles, lattice volumes, pion and nucleon masses, as well as the number of configurations, 
of measurements and of available source-sink separations $t_s$, are listed in Table~\ref{tab:ensembles}.
All lattices used in this study have a fairly large volume, which is indicated by $M_\pi L\gtrsim 4$. 

\subsection{Method}

\begin{figure}[!hb]    
\centering
\includegraphics[width=0.58\columnwidth]{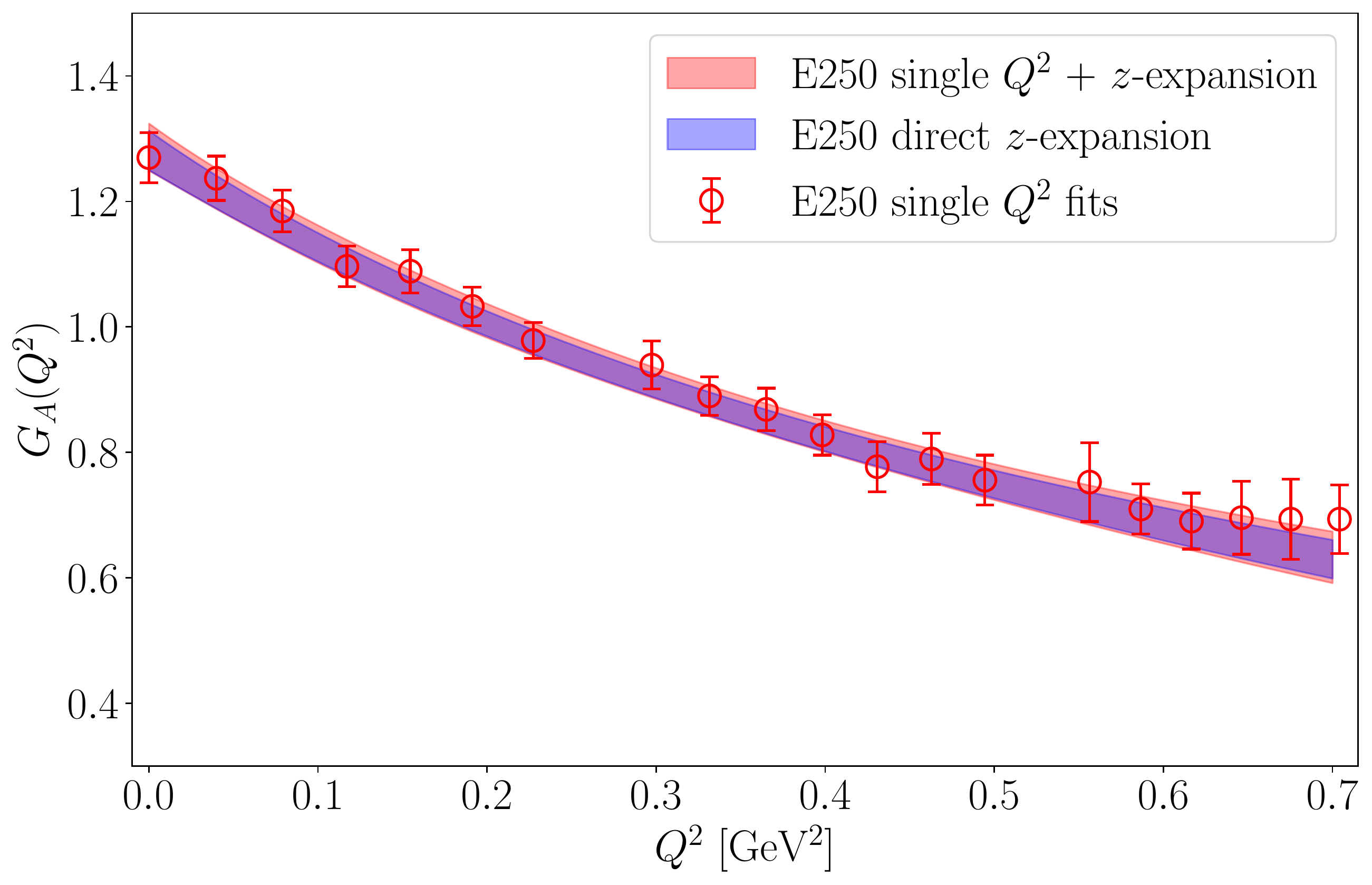}
\caption{Illustration of the new method. We plot the results of fits to single $Q^2$
data (red data points) and compare them to the result of the immediate $z$-expansion (the blue error
band) on ensemble E250 using $t_s^{\textrm{min}}=12$. The red error band is a $z$-expansion fit to the
single $Q^2$ data points (including correlations), whereas the blue error band shows the fit where the
$z$-expansion is directly incorporated into the summation method (Eq.~\eqref{eq:summation}). Here
$t_s^{\textrm{min}} = 12$. The agreement is good, and the immediate $z$-expansion provides readily a
model-independent parameterization of the shape of the form factor.}
\label{fig:E250_singleQ2_vs_zfit}
\end{figure}

In our analysis, we incorporate the $z$-expansion, which parameterizes the $Q^2$ dependence of the form factor,
directly into the summation method. This can also be done in two separate steps, first using the summation method
to get the value of the form factor at a given $Q^2$, keeping track of the correlation
between the data at different $Q^2$ values, and then parameterizing the $Q^2$ dependence of the form factor
using a $z$-expansion. The two methods should obviously give compatible results, which we find to be the case.
This is illustrated in Fig.~\ref{fig:E250_singleQ2_vs_zfit} on ensemble E250. The first method leads to larger
correlation matrices, which is why we have to damp the off-diagonal elements in some cases (for matrices larger
than $70\time 70$) by 0.5\%\ldots1.5\%. However, the one-step fits are very stable and robust, and the damping
of off-diagonal correlations essentially only affects the $\chi^2$ of the fit. This is our preferred method, as
it gives readily a parameterization of the shape of the form factor. The results of these fits are tabulated
in Table~\ref{tab:coefficients} ensemble-by-ensemble.

\subsection{Extrapolation to physical point and FSE}

\begin{center}
\begin{figure}[!hb]    
\includegraphics[width=0.4\columnwidth]{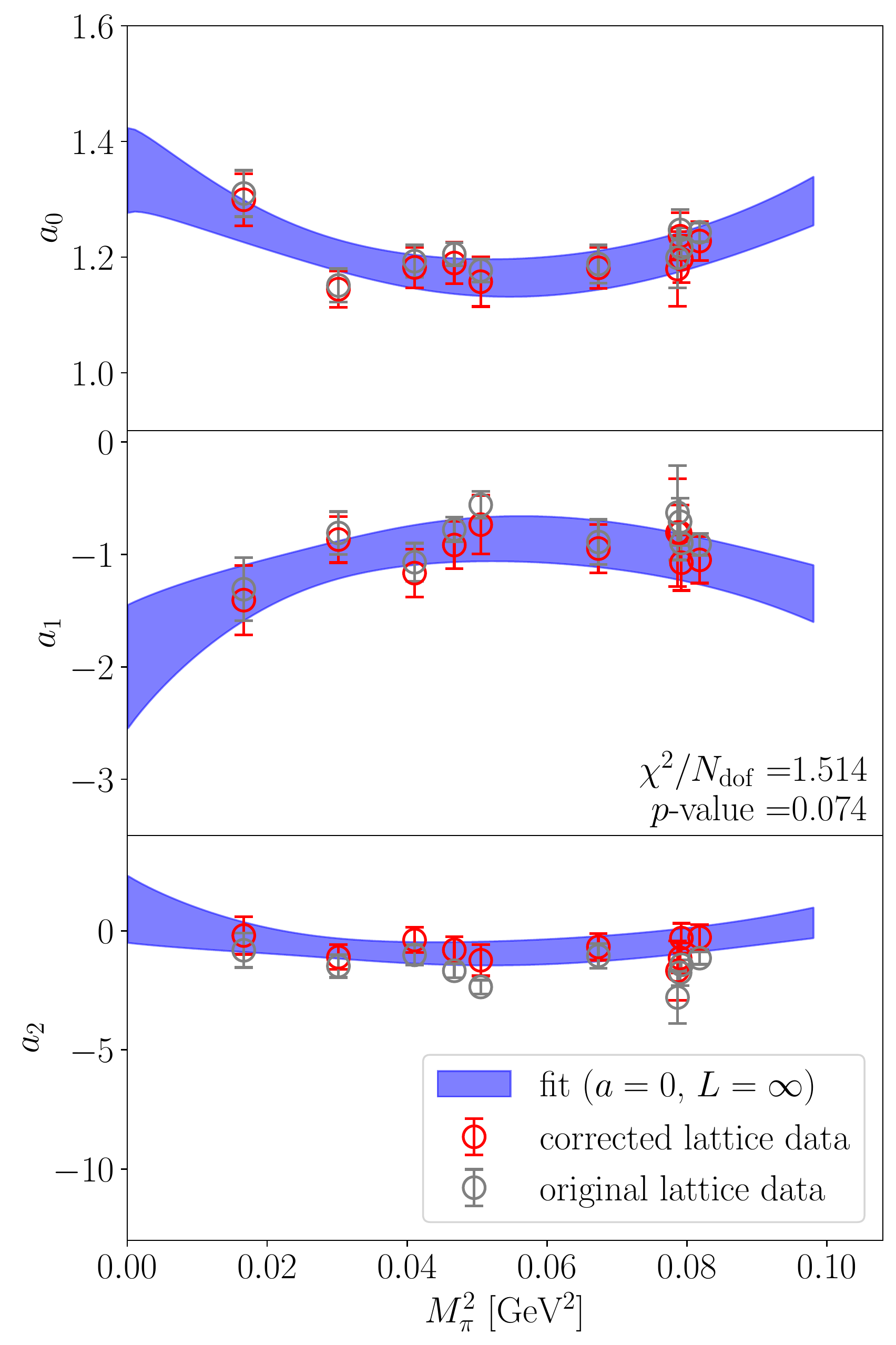}\hspace*{16mm}
\includegraphics[width=0.4\columnwidth]{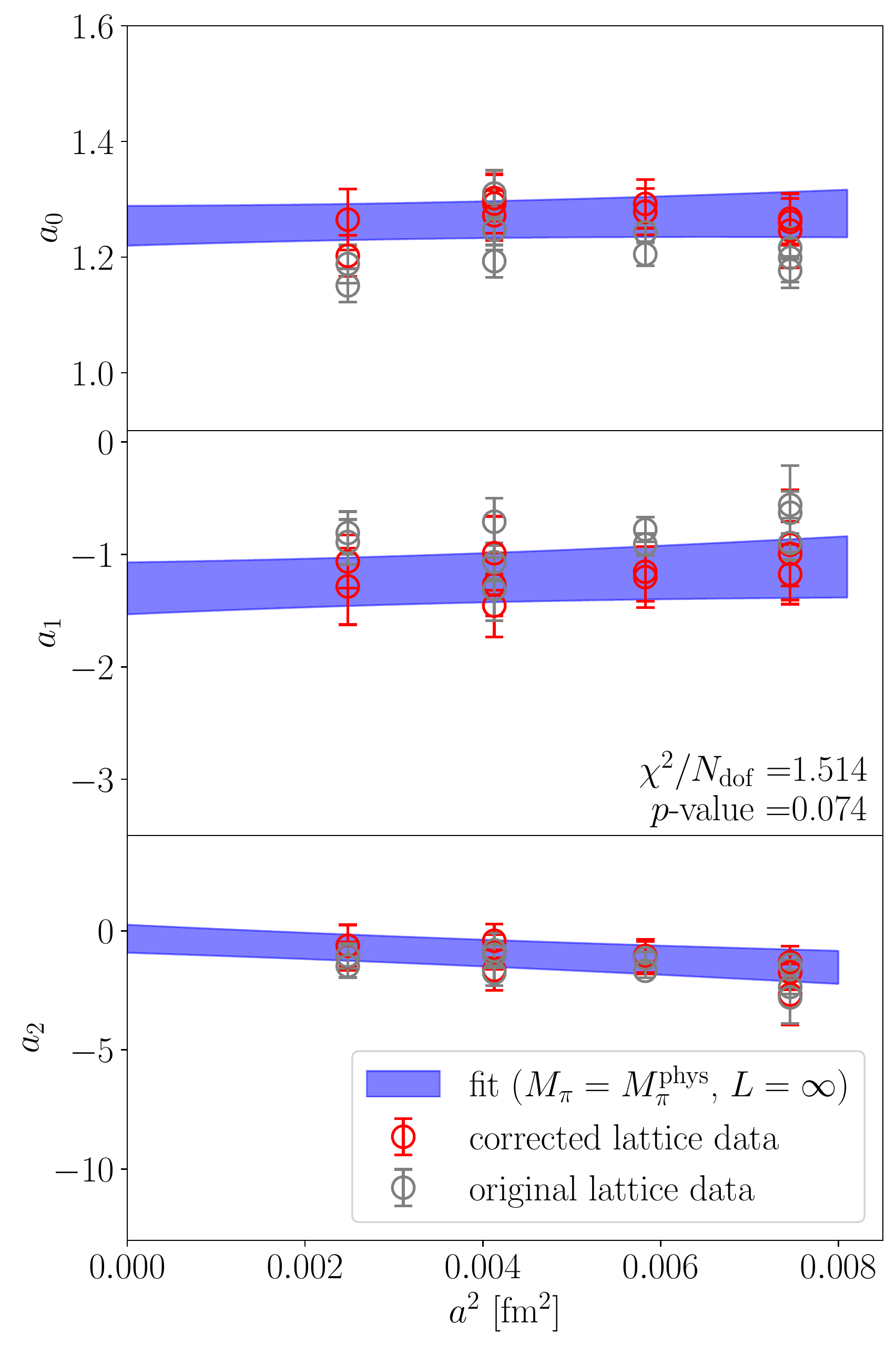}
\caption{Chiral and continuum extrapolations using fit ansatz 3 (without the FSE term)
  and a pion mass cut of 300 MeV. The red circles show the corrected lattice data at infinite volume, and with
  zero lattice spacing or physical pion mass respectively in the left and right columns, whereas the grey data points are uncorrected.}
\label{fig:extrapolation-mpicut0.3-ansatz3}
\end{figure}
\end{center}

\begin{center}
\begin{figure}[!ht]    
\includegraphics[width=0.39\columnwidth]{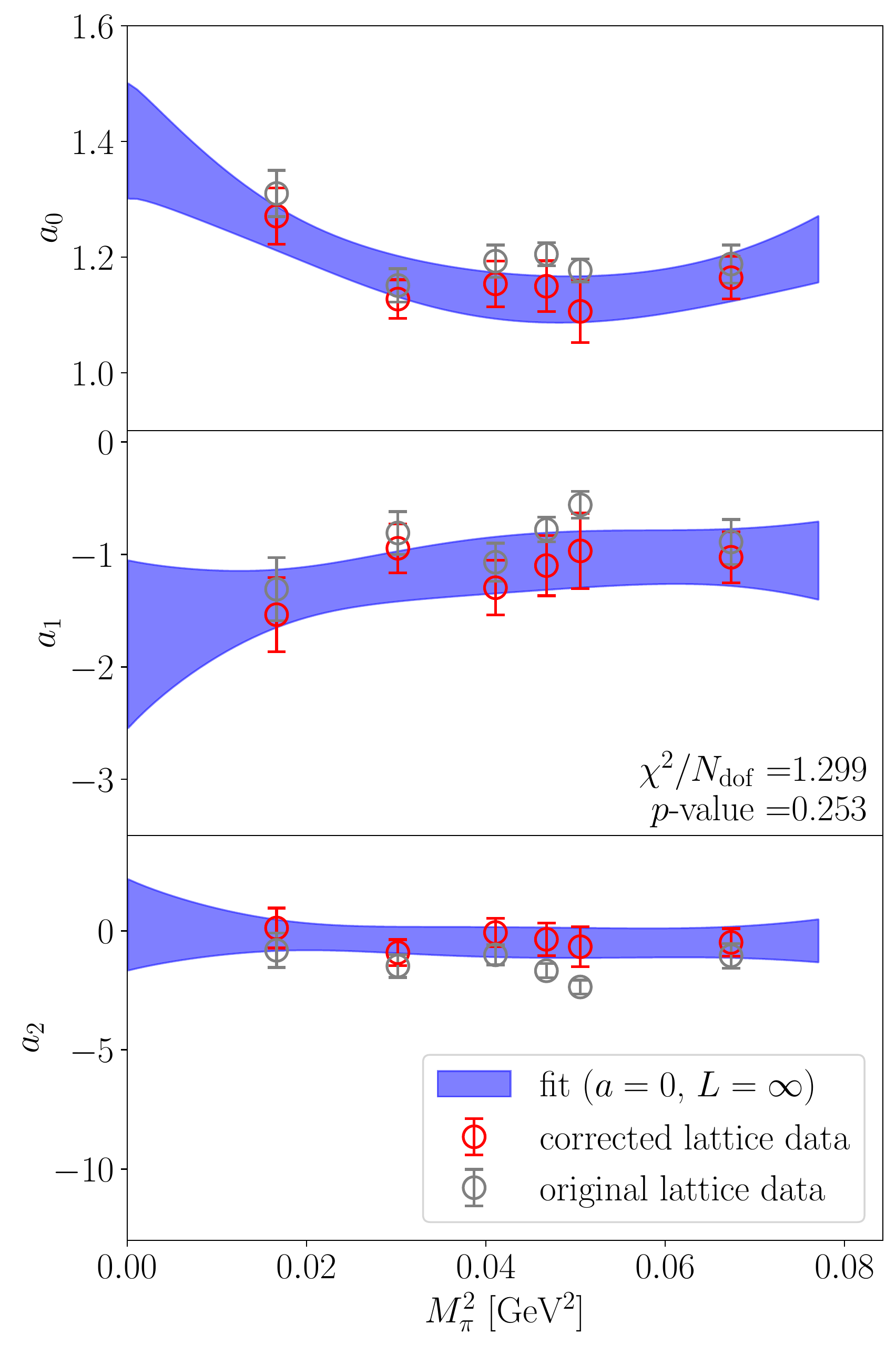}\hspace*{8mm}
\includegraphics[width=0.39\columnwidth]{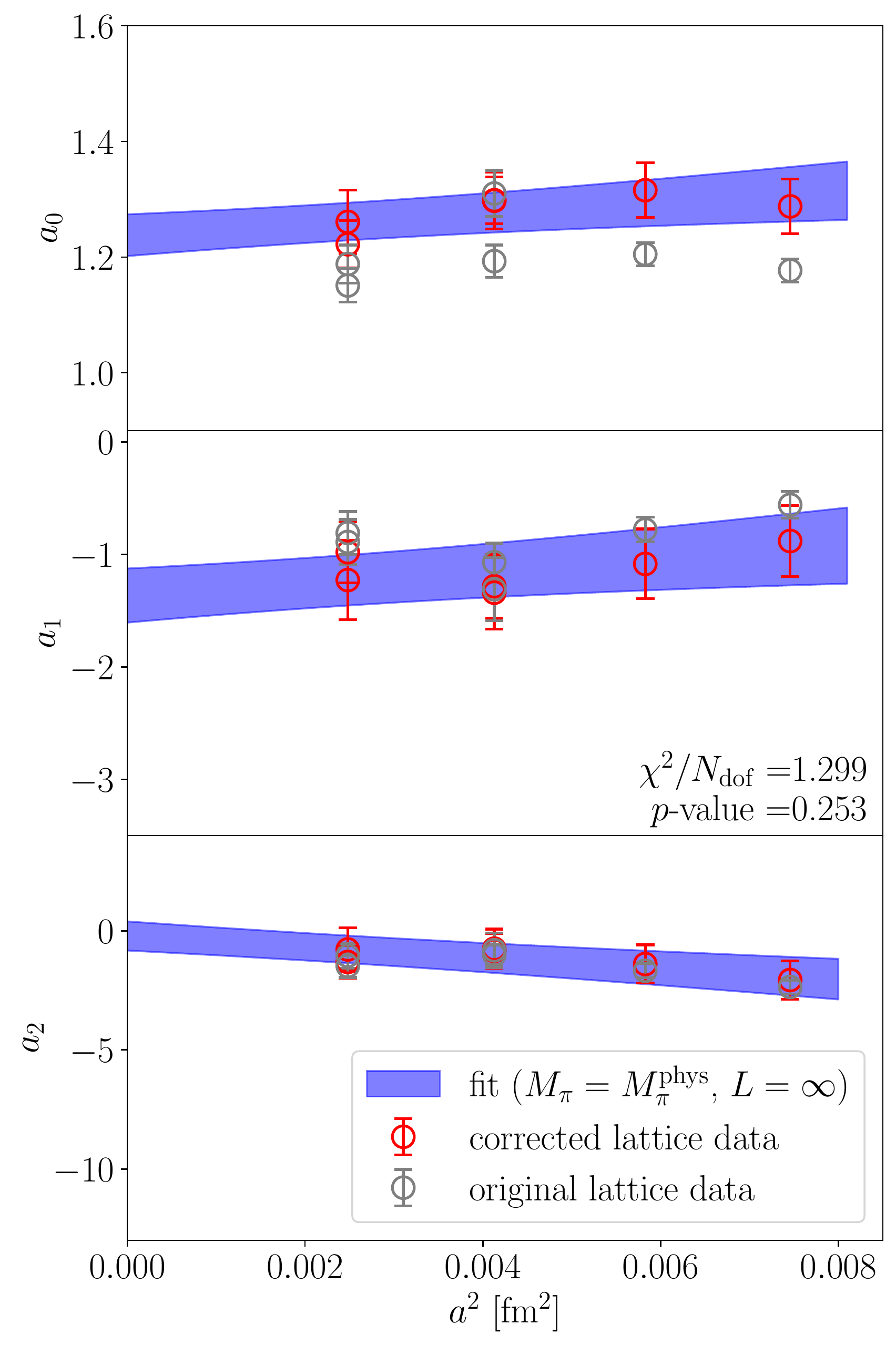}\\
\includegraphics[width=0.39\columnwidth]{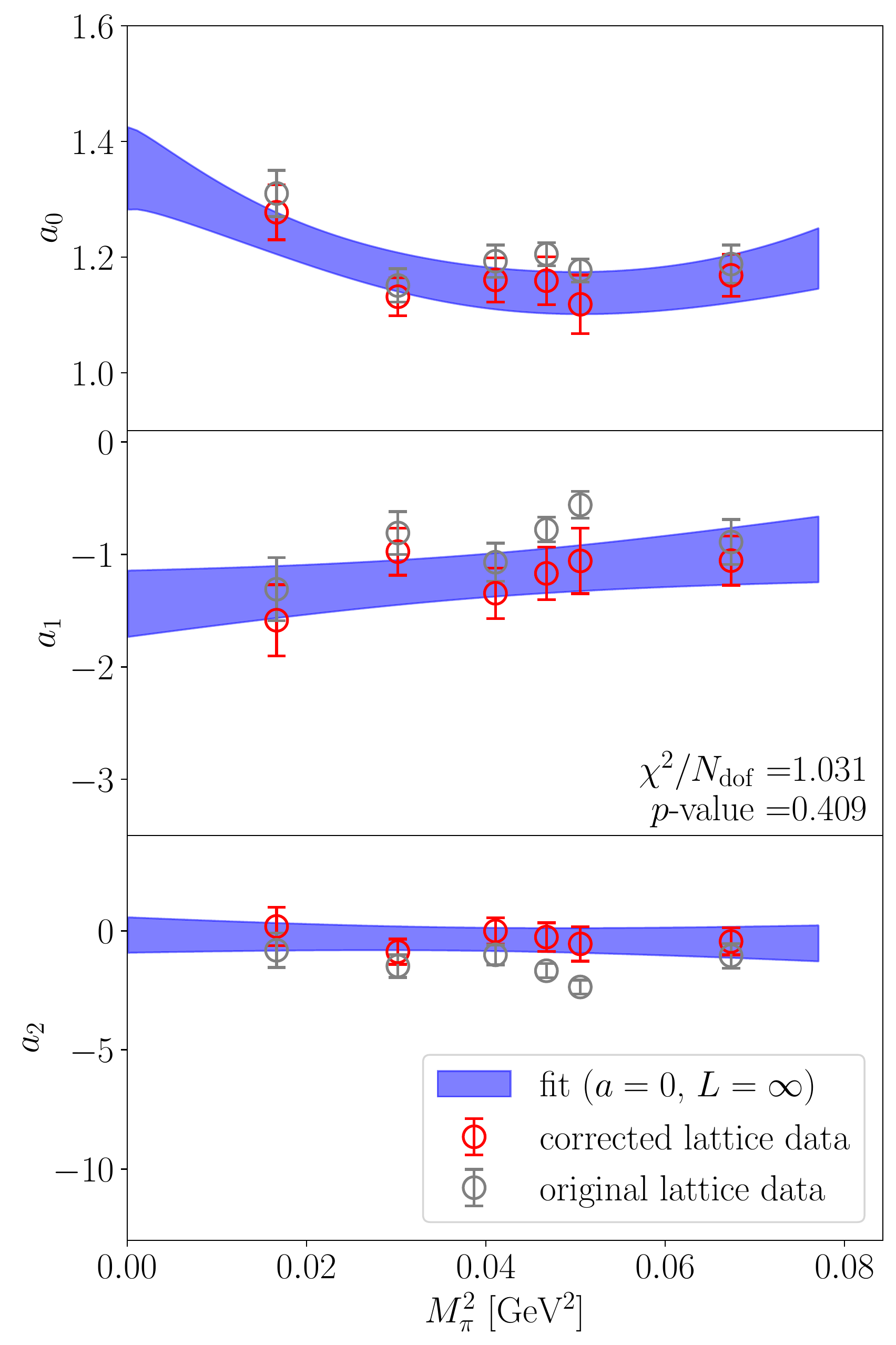}\hspace*{8mm}
\includegraphics[width=0.39\columnwidth]{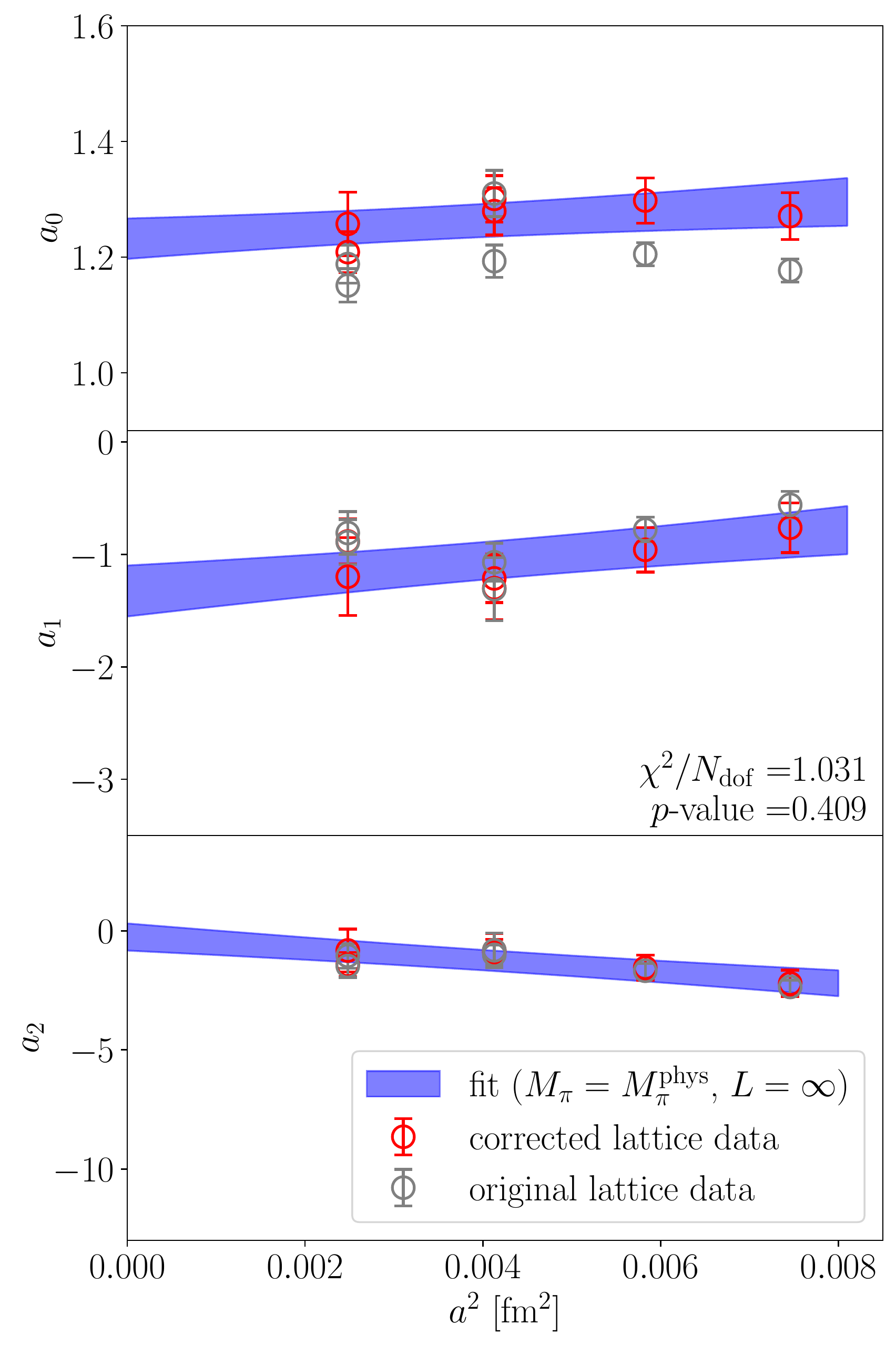}
\caption{Extrapolation in pion mass and lattice spacing using fit ansatz 3 (top panel) and fit
  ansatz 2 (lower panel) with a pion mass cut of 265 MeV. No FSE term was included in these fits.
  The red circles show the corrected lattice data at infinite volume, and with zero lattice spacing or
  physical pion mass respectively in the left and right columns, whereas the grey data points are uncorrected.}
\label{fig:extrapolation-mpicut0.265-ansatz2n3}
\end{figure}
\end{center}

\begin{figure}[!ht]    
\centering
\includegraphics[width=0.42\columnwidth]{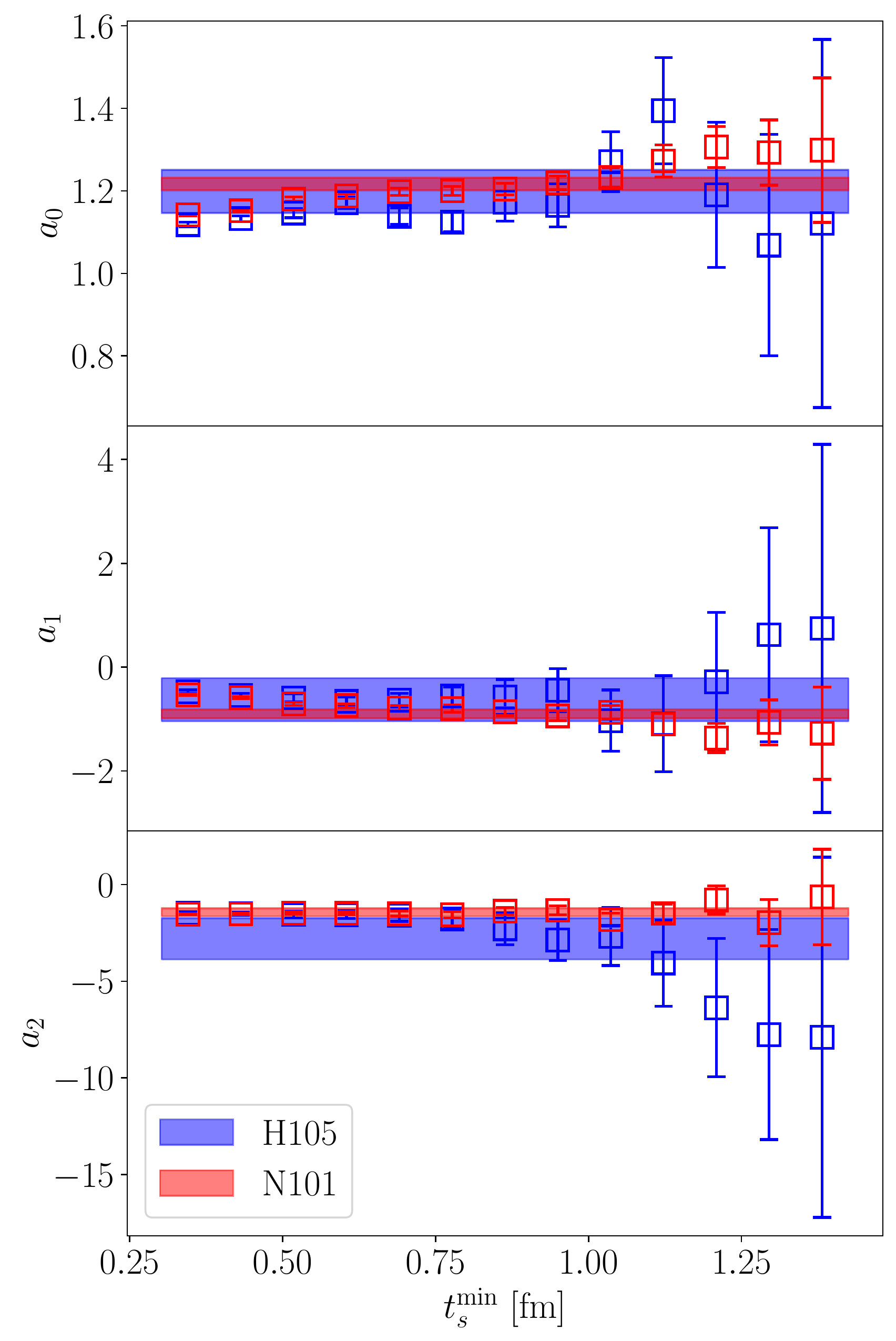}
\caption{Direct comparison of the data as a function of $t_{s}^{\mathrm{min}}$ and the 'smooth window'
average of Eq.~\eqref{eq:smoothwindow} on two of the ensembles, H105 and N101. These ensembles have the
same lattice spacing and quark masses, but differ by their volume. The
data points show good agreement, which indicates that any finite size effects are small.}
\label{fig:compare_H105_N101}
\end{figure}

\begin{center}
\begin{figure}[!hb]    
\includegraphics[width=0.49\columnwidth]{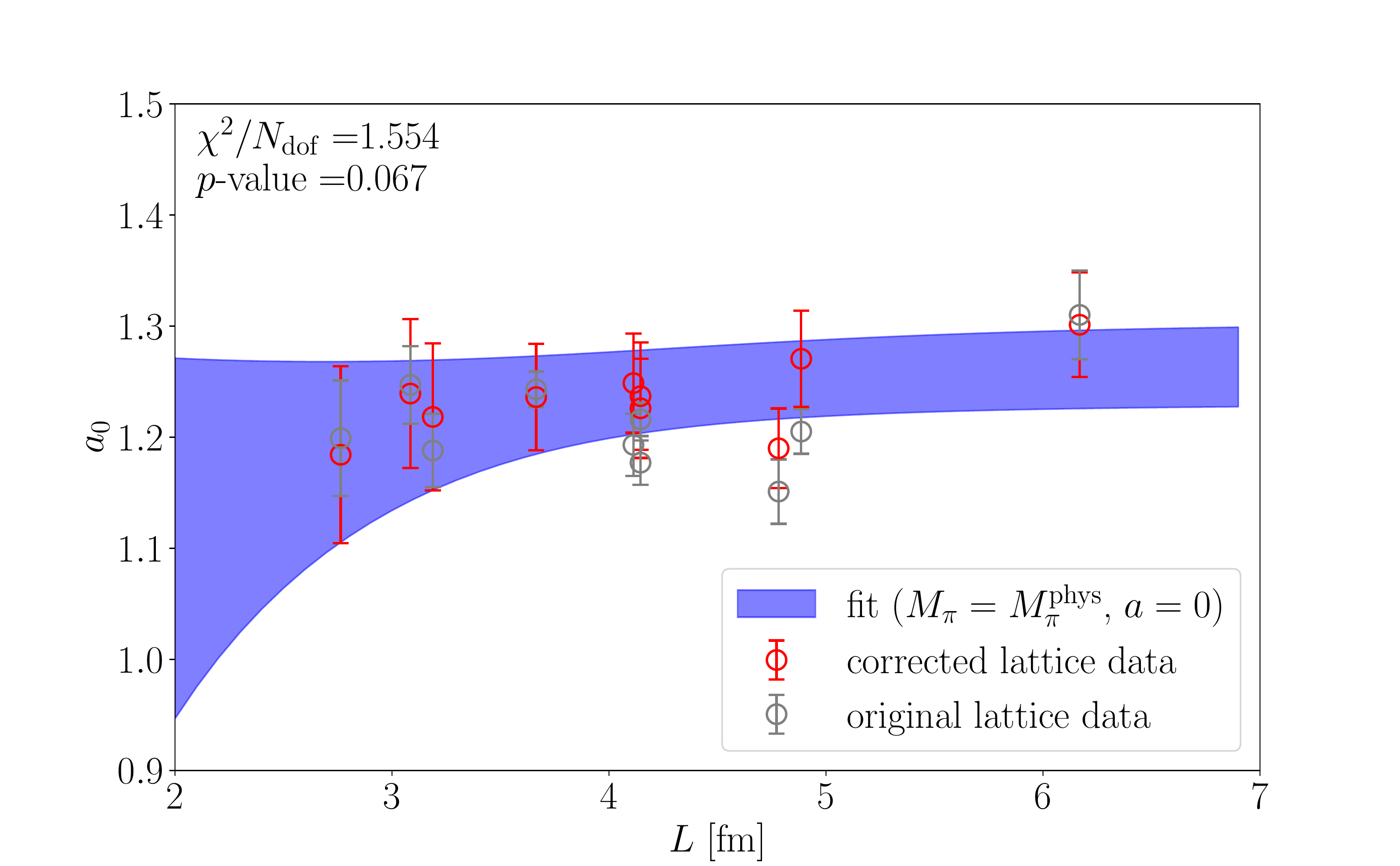}
\includegraphics[width=0.49\columnwidth]{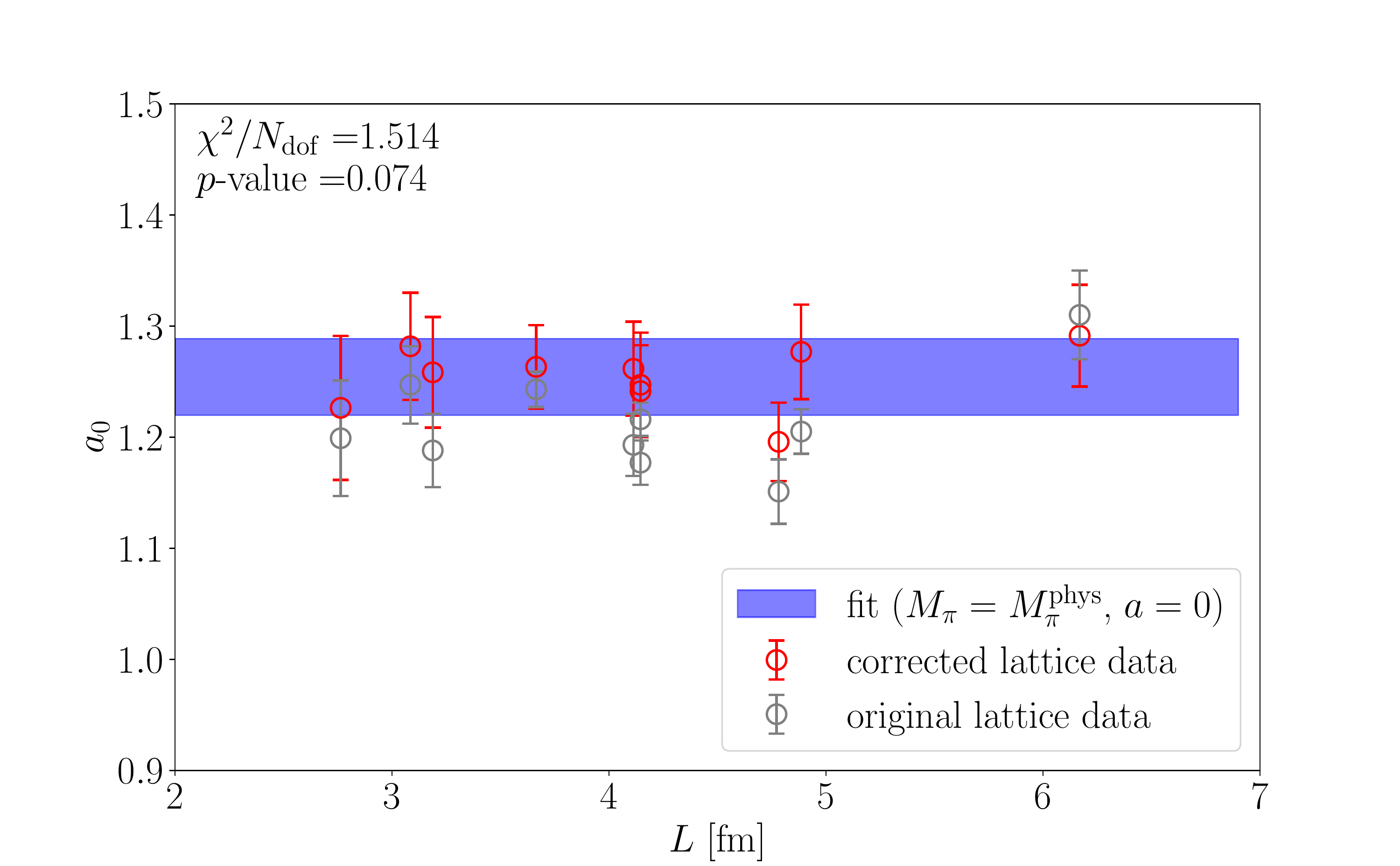}
\caption{Finite size effects: coefficient $a_0$ (the axial charge) from fit ansatz 3
  with a pion mass cut of 300 MeV, with and without the FSE term (Eq.~\eqref{eq:FSE}).
  The two fits are equally good, and are both included in the AIC average.}
\label{fig:ff_extrapolation_FSE}
\end{figure}
\end{center}

We include global fits with three different ansätze for the chiral behaviour of the form factor and several
pion mass cuts in our final AIC average. We also take into account finite volume corrections by including
a volume-dependent term (Eq.~\eqref{eq:FSE} in the main text) in some of our fits. We show examples of these
global fits in Figs.~\ref{fig:extrapolation-mpicut0.3-ansatz3}, \ref{fig:extrapolation-mpicut0.265-ansatz2n3} and
\ref{fig:ff_extrapolation_FSE}. Fig.~\ref{fig:extrapolation-mpicut0.265-ansatz2n3} highlights the difference
between ansatz 2 and ansatz 3, whereas comparing Fig.~\ref{fig:extrapolation-mpicut0.3-ansatz3} and
the top panel of Fig.~\ref{fig:extrapolation-mpicut0.265-ansatz2n3} shows ansatz 3 with different pion
mass cuts (300 MeV and 265 MeV, respectively). Fig.~\ref{fig:ff_extrapolation_FSE} compares a selected fit,
ansatz 3 with a pion mass cut of 300 MeV, with and without the FSE term. At present statistics, the effect
of adding the FSE term to the fit is almost negligible. Doing a more direct comparison of finite volume effects
by looking at two ensembles, N101 and H105, which differ only by their volume, confirms this. We show the
results of the $z$-expansion fits on these two ensembles as a function of $t_s^{\textrm{min}}$ in
Fig.~\ref{fig:compare_H105_N101}. We find that the coefficients $a_i$ are consistent between the two ensembles,
and observe no significant finite size effects.

\subsection{Akaike (AIC) model average}

In this section, we give more details of the final step of the analysis, the AIC model average. As
discussed in section~\ref{sec:lattice}, we take systematic errors into account by performing an
Akaike-information-criterion based average over a set of chiral, continuum and infinite-volume
extrapolations. We choose the weight~\cite{Borsanyi:2020mff}
\begin{equation*}
  w^{\textrm{AIC}}_k=N\mathrm{e}^{-\frac{1}{2}\left(\chi^2_k+2n_{\textrm{par},k}-n_{\textrm{data},k}\right)},
\end{equation*}
where the $\chi^2_k$, the number of fit parameters $n_{\textrm{par},k}$ and the number of data points $n_{\textrm{data},k}$
describe the $k$-th global fit. The first two terms in the exponent correspond to the standard AIC, and the last term
is introduced to take into account fits with different number of data points, i.e. fits with different cuts
in pion mass or lattice spacing. The weights are normalized so that $\sum_i w_i = 1$.

The weights $w^{\textrm{AIC}}_i$ are interpreted as propabilities, and the analyses follow a normal (Gaussian)
distribution $N(a_i;m_k,\sigma_k)$ with a central value $m_k$ and a standard deviation $\sigma_k$ for the quantity
$a_i$. $m_k$ and $\sigma_k$ are the jackknife average and the jackknife error in the $k$-th analysis. A
joint distribution function can then be defined as
\begin{equation*}
\sum_k w^{\textrm{AIC}}_kN(a_i;m_k,\sigma_k),
\end{equation*}
which includes both statistical and systematic uncertainties. The corresponding cumulative distribution function
reads
\begin{equation*}
P(a_i)=\int_{-\infty}^{a_i}\mathrm{d}a_i' \sum_k w^{\textrm{AIC}}_kN(a_i';m_k,\sigma_k).
\end{equation*}
The median of the CDF gives the central value of $a_i$ and its total error is given by the 16\% and 84\%
percentiles of the CDF. Noticing that scaling $\sigma_k$ by a factor of $\sqrt{\lambda}$ scales the
statistical error by $\sqrt{\lambda}$, but does not scale the systematic error, using $\lambda=1$ and $\lambda=2$
allows us to calculate the break-up of the total uncertainty into statistical and systematic parts.

\begin{center}
\begin{figure}[!ht]    
\centering
\includegraphics[width=0.6\textwidth]{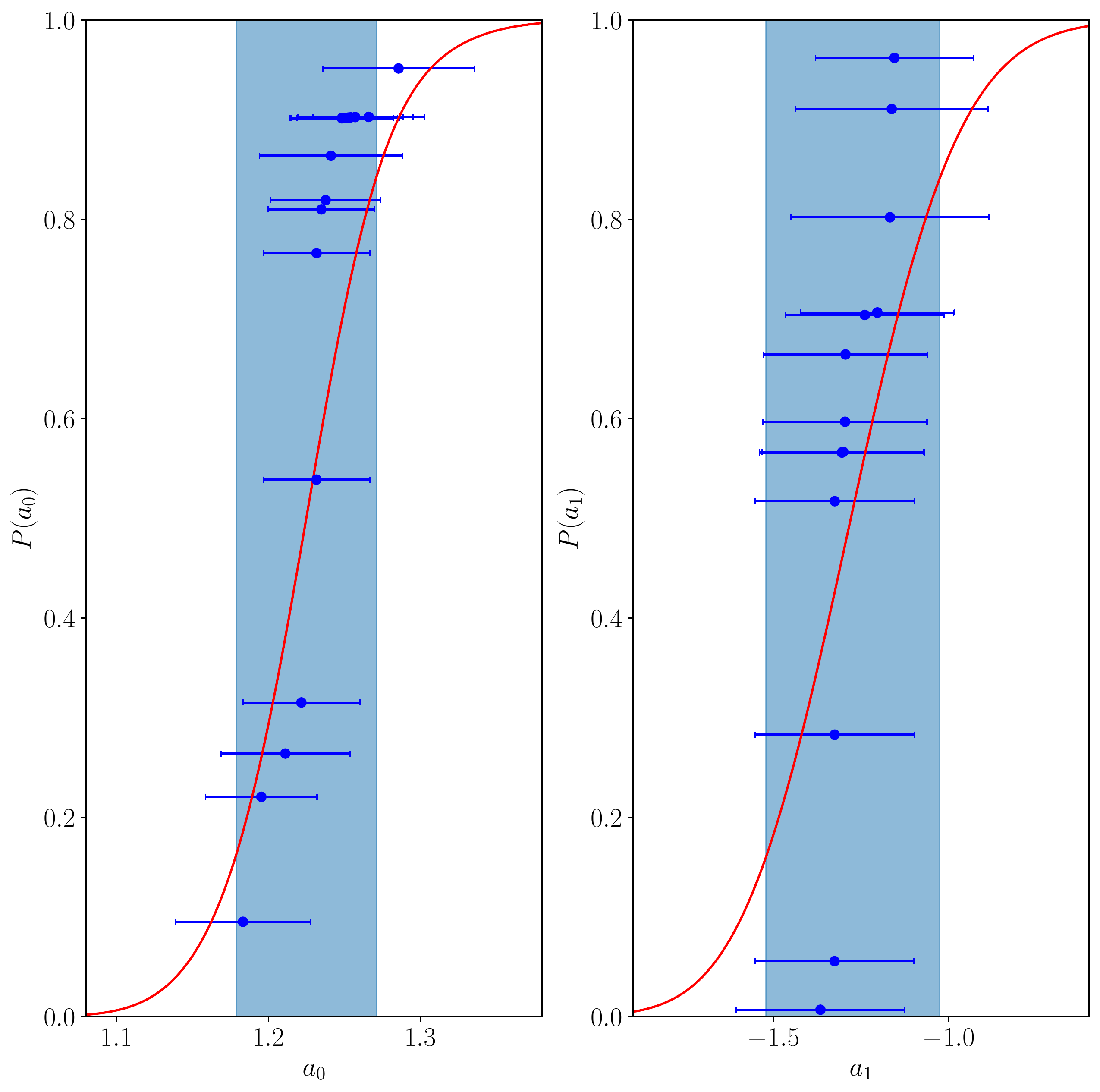}\\
\includegraphics[width=0.6\textwidth]{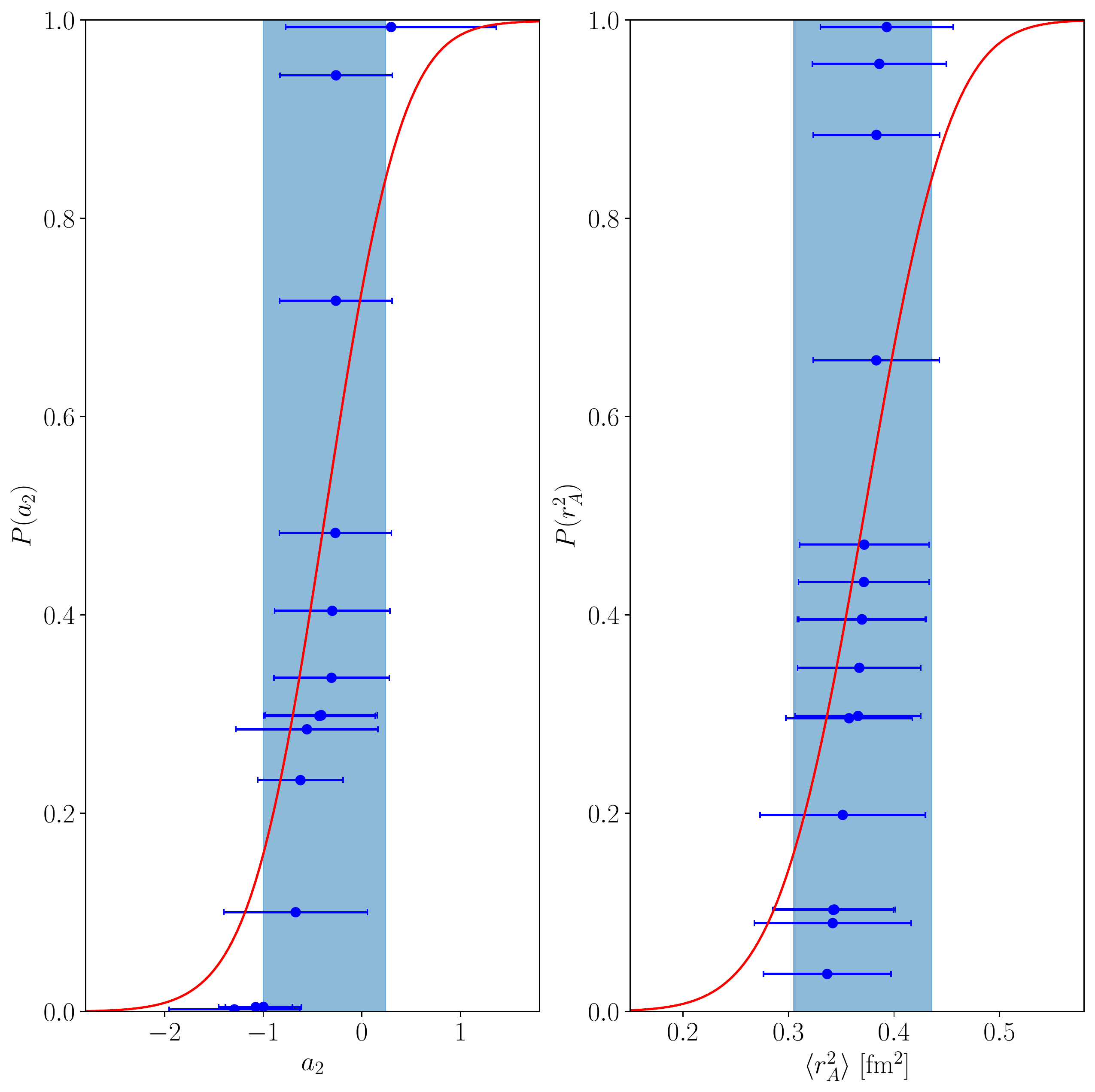}
\caption{AIC average and the corresponding cumulative distribution function for all coefficients
$a_0$, $a_1$, $a_2$, and the mean square radius $\langle r_A^2\rangle$ (in fm$^2$ units).}
\label{fig:a0_rA_AIC}
\end{figure}
\end{center}

In Fig.~\ref{fig:a0_rA_AIC}, we show the AIC averages and the corresponding cumulative distributions for all
coefficients $a_i$ as well as for the mean square radius $\langle r_A^2\rangle$. These are all well-behaved
and contain no outliers. The data points are individual analyses, or fits, that give a good
description of the data with a $p\,$-value better than 5\%. These are the analyses that enter the AIC procedure.
The error band shows the AIC average with the total (statistical and systematic) uncertainty.

\onecolumngrid

\begin{table}[!ht]
 \centering
  \begin{tabular}{lcccccccccr}
   \hline\hline
   ID  & $\beta$ & $T/a$ & $L/a$ & $M_\pi~[\mev]$ & $M_\pi L$ & $M_N~[\gev]$ & $N_\mathrm{conf}$ & $N_\mathrm{meas}$ & $t_s~[\fm]$ & $N_{t_s}$ \\
   \hline\hline
   H102 & 3.40 &  96 & 32 & 354 & 4.96 & 1.103 & 2005 &  32080 & 0.35..1.47 & 14 \\ 
   H105 & 3.40 &  96 & 32 & 280 & 3.93 & 1.045 & 1027 &  49296 & 0.35..1.47 & 14 \\ 
   C101 & 3.40 &  96 & 48 & 225 & 4.73 & 0.980 & 2000 &  64000 & 0.35..1.47 & 14 \\ 
   N101 & 3.40 & 128 & 48 & 281 & 5.91 & 1.030 & 1596 &  51072 & 0.35..1.47 & 14 \\ 
   \hline                                                                    
   S400 & 3.46 & 128 & 32 & 350 & 4.33 & 1.130 & 2873 &  45968 & 0.31..1.53 &  9 \\ 
   N451 & 3.46 & 128 & 48 & 286 & 5.31 & 1.045 & 1011 & 129408 & 0.31..1.53 &  9 \\ 
   D450 & 3.46 & 128 & 64 & 216 & 5.35 & 0.978 &  500 &  64000 & 0.31..1.53 & 17 \\ 
   \hline                                                                    
   N203 & 3.55 & 128 & 48 & 346 & 5.41 & 1.112 & 1543 &  24688 & 0.26..1.41 & 10 \\ 
   N200 & 3.55 & 128 & 48 & 281 & 4.39 & 1.063 & 1712 &  20544 & 0.26..1.41 & 10 \\ 
   D200 & 3.55 & 128 & 64 & 203 & 4.22 & 0.966 & 2000 &  64000 & 0.26..1.41 & 10 \\ 
   E250 & 3.55 & 192 & 96 & 129 & 4.04 & 0.928 &  400 & 102400 & 0.26..1.41 & 10 \\ 
   \hline                                                                    
   N302 & 3.70 & 128 & 48 & 348 & 4.22 & 1.146 & 2201 &  35216 & 0.20..1.40 & 13 \\ 
   J303 & 3.70 & 192 & 64 & 260 & 4.19 & 1.048 & 1073 &  17168 & 0.20..1.40 & 13 \\ 
   E300 & 3.70 & 192 & 96 & 174 & 4.21 & 0.962 &  570 &  18240 & 0.20..1.40 & 13 \\ 
   \hline\hline
   \vspace*{0.1cm}
  \end{tabular}
  \caption{Overview of ensembles used in this study. The values $\beta=3.40$, $3.46$, $3.55$ and $3.70$
  correspond to lattice spacings $a\approx 0.086$, $0.076$, $0.064$ and $0.05~\fm$, respectively. Columns
  $T/a$ and $L/a$ give the temporal and spatial size of the lattice, and $M_\pi$ and $M_N$ are the pion
  and nucleon masses. $N_\mathrm{conf}$ is the number of configurations used for each ensemble, and
  in column $N_\mathrm{meas}$ we list the maximum number of measurements done per $t_s$ at largest source-sink
  separations ($> 1.0~\fm$). To keep the signal-to-noise ratio as a function of $t_s$ close to constant,
  the number of measurements is reduced by a factor of two in steps of $\Delta t_s\approx 0.2~\fm$ for
  $t_s < 1.0~\fm$. $N_{t_s}$ is the number of available source-sink separations in the range listed in column
  $t_s$. However, in this study we restrict the smallest value of $t_s$ included in the summation method,
  $t_s^{\textrm{min}}$, to be in the range $0.8..1.0~\fm$ using the smooth window (see Eq.~\eqref{eq:smoothwindow}).}
 \label{tab:ensembles}
\end{table}

\begin{table}[!hb]
 \centering
  \begin{tabular}{lcccccc}
   \hline\hline
   ID  & $a_0$ & $a_1$ & $a_2$ & $\rho_{a_0,a_1}$ & $\rho_{a_0,a_2}$ & $\rho_{a_1,a_2}$ \\
   \hline
 C101 & $1.177(20)$ & $-0.56(12)$  & $-2.36(29)$ & $-0.59707$ & $0.30568$ & $-0.85694$ \\
 D200 & $1.193(28)$ & $-1.07(17)$  & $-1.01(42)$ & $-0.58273$ & $0.35465$ & $-0.87957$ \\
 D450 & $1.205(20)$ & $-0.78(11)$  & $-1.68(30)$ & $-0.52813$ & $0.13464$ & $-0.76168$ \\
 E250 & $1.310(40)$ & $-1.31(28)$  & $-0.82(72)$ & $-0.61384$ & $0.29885$ & $-0.87065$ \\
 E300 & $1.151(29)$ & $-0.81(19)$  & $-1.48(48)$ & $-0.58834$ & $0.31474$ & $-0.88040$ \\
 H102 & $1.157(16)$ & $-0.55(11)$  & $-2.01(33)$ & $-0.39306$ & $0.12100$ & $-0.89438$ \\
 H105 & $1.199(52)$ & $-0.63(42)$  & $-2.8(1.1)$ & $-0.55469$ & $0.33040$ & $-0.93019$ \\
 J303 & $1.188(33)$ & $-0.89(20)$  & $-1.05(52)$ & $-0.59229$ & $0.24273$ & $-0.84407$ \\
 N101 & $1.216(15)$ & $-0.899(86)$ & $-1.43(21)$ & $-0.55315$ & $0.22582$ & $-0.81047$ \\
 N200 & $1.247(35)$ & $-0.71(21)$  & $-1.76(54)$ & $-0.53414$ & $0.24793$ & $-0.86773$ \\
 N203 & $1.123(23)$ & $-0.66(13)$  & $-1.54(35)$ & $-0.48820$ & $0.18570$ & $-0.83382$ \\
 N302 & $1.164(34)$ & $-0.64(26)$  & $-2.36(70)$ & $-0.48779$ & $0.21775$ & $-0.91619$ \\
 N451 & $1.243(16)$ & $-0.912(99)$ & $-1.15(26)$ & $-0.54656$ & $0.27504$ & $-0.84718$ \\
 S400 & $1.178(23)$ & $-0.47(18)$  & $-1.99(51)$ & $-0.44271$ & $0.18507$ & $-0.91340$ \\
   \hline\hline
   \vspace*{0.1cm}
  \end{tabular}
  \caption{Our results for the coefficients $a_0$, $a_1$, $a_2$ of the $z$-expansion for each ensemble,
  as well as their correlations
  $\rho_{a_i,a_j}= (\langle a_i a_j\rangle - \langle a_i\rangle \langle a_j\rangle)/(\sqrt{\langle a_i^2\rangle - \langle a_i\rangle^2}\sqrt{\langle a_j^2\rangle -\langle a_j\rangle^2})$.
  These are smooth window averages (see Eq.~\eqref{eq:smoothwindow}) of
  $z$-expansion fits to the sum $S(\vec q,t_s)$ in Eq.~\eqref{eq:summation} using different $t_s^{\textrm{min}}$.}
 \label{tab:coefficients}
\end{table}

\clearpage
